\documentclass[aps,prl,reprint,groupedaddress]{revtex4-2}

\usepackage[utf8]{inputenc}
\usepackage{physics}
\usepackage{amsmath,amsthm,amssymb}
\usepackage{mathrsfs}
\usepackage{graphicx}
\usepackage{dcolumn}
\usepackage{bm}
\usepackage{siunitx}
\usepackage{caption}
\usepackage{subcaption}
\usepackage[hidelinks]{hyperref}

\begin{document}
\title{Asymptotic Throat: The Geometric Inevitability of Regular Black Holes}
	
\author{Yi-Bo Liang}
 \email[]{liangyibo@stu.xjtu.edu.cn}
\author{Hong-Rong Li}
\email[Corresponding author: ]{hrli@xjtu.edu.cn}
\affiliation{School of Physics, Xi’an Jiaotong University, Xi’an 710049, China}
	
\date{April 13 2026}
\begin{abstract}
	We reveal that a fundamental minimal length naturally replaces the Schwarzschild singularity with future infinity, formalizing the ``asymptotic throat'' as a geometric inevitability.
	This scheme avoids the topology changes, multiple horizons, and universe towers characteristic of existing regular black hole models.
	We establish a general regularization framework, construct explicit examples with their physical sources, and show that the surface gravity and Hawking temperature remain unaltered.
	The asymptotic throat thus provides a pristine classical bedrock for future quantum gravity investigations.

\end{abstract}
	
\maketitle
	\textit{Introduction.}---
	Black holes stand as both a cornerstone of Einstein’s theory of gravity and a firmly established astrophysical reality.
	The Schwarzschild solution \cite{schwarzschild1916gravitationsfeld} provides the foundational framework for understanding these objects; however, its interior harbors a spacetime singularity \cite{penrose1965gravitational,hawking1967occurrence,hawking1970singularities}.
	This singularity signals a catastrophic breakdown of classical predictability, preventing a smooth, deterministic global extension of the manifold.
	From a fundamental perspective, a well-defined physical spacetime should be free of such pathological boundaries.
	
	Contemporary efforts to resolve this singularity have converged on the concept of regular black holes---nonsingular, horizoned objects that eliminate curvature divergences, provide a geodesically complete spacetime, and offer a well-defined background for semiclassical and quantum field theory.
	These models predominantly fall into two distinct paradigms.
	The first involves a regular center, typically realized as a de Sitter or Minkowski core, which inevitably necessitates a topology change \cite{bardeen1968non,ayon1998regular,ayon2000bardeen,bronnikov2001regular,burinskii2002new,fan2016construction,bronnikov2023regular}.
	The second explores the black bounce mechanism, wherein a spacelike transition surface inside the horizon connects the trapped region of one universe to an anti-trapped region in another \cite{simpson2019black, lobo2021novel, bronnikov2022black, bronnikov2022field, canate2022black}.
	While successful in removing the singularity, both schemes invoke substantial modifications to the global spacetime structure.
	
	In this Letter, we introduce a fundamentally distinct and geometrically inevitable mechanism for singularity resolution.
	The underlying premise is minimal: since the discretization of spacetime is expected to forbid the compression of matter beyond the Planck volume \cite{rovelli2004quantum}, gravitational collapse in a Schwarzschild geometry must terminate at a finite minimal radius $L$.
	Due to the well-known signature exchange of temporal and radial coordinates inside the horizon, constant-radius hypersurfaces are spacelike.
	Consequently, without invoking any additional exotic physics, this minimal radius is naturally identified as future infinity.
	This framework formalizes the concept of the asymptotic throat \cite{liang2025globally}---an unattainable minimal-radius structure that acts as a causal boundary---elevating it from a heuristic idea to a rigorous geometric construction.
	
	Based on influential models that suggested black holes may transition into traversable wormholes and vice versa \cite{hayward1999dynamic, shinkai2002fate, hayward2009wormhole, hayward2006formation, simpson2019vaidya, lobo2020dynamic, yang2021trapping}, the asymptotic throat concept (a null-type ideal boundary) was recently introduced and analyzed in \cite{liang2025globally}.
	In this context, it is worth noting that earlier studies had already hinted at related geometric possibilities: some mentioned the possibility of constant-radius boundaries \cite{bronnikov2006regular}, while others presented a spacelike-type counterpart termed the asymptotic hidden wormhole \cite{carballo2020geodesically}.
	The present work reveals the theoretical possibility of the existence of such regularized spacetimes; a concrete constructive scheme, however, has remained elusive.
	We will adopt the asymptotic throat concept for simplicity and precision, since the existence of such a minimal radius means that it is always possible to find a spacelike slice that has a throat with radius tending to the minimal radius. 
	
	The asymptotic throat paradigm offers distinct conceptual and structural advantages over existing schemes: 1. Geometric Minimalism: It does not require the multiple horizons or topology changes mandated by regular-center models; 2. Causal Simplicity: It avoids the infinite-universe tower characteristic of black bounce geometries; 3. Inevitability: The minimal throat emerges as a natural and unavoidable consequence of imposing a fundamental length cutoff.
	Specifically, no exact scheme to date has simultaneously achieved: (i) reduction to the Schwarzschild metric in the appropriate weak-field limit; (ii) preservation of global hyperbolicity, asymptotic flatness, and a standard bifurcate Killing horizon.
	The asymptotic throat framework presented in this Letter fulfills both these criteria.
	It thus provides the first concrete realization of a regular black hole that simultaneously satisfies these fundamental requirements.
	
	In this work, we establish a general framework to explicitly regularize the Schwarzschild spacetime.
	As concrete proofs of concept, we present two explicit examples---yielding null-type and spacelike-type future infinities, respectively---and identify their underlying physical sources within a scalar-electromagnetic system.
	Furthermore, we derive the corresponding causal diagrams and confirm that the surface gravity remain unaltered.
	This regularization scheme is applicable to a wide class of black hole geometries, providing a pristine classical bedrock for future investigations into the quantum nature of black hole interiors.
	
	\textit{General Framework.}---
	For a maximally extended Schwarzschild black hole, its metric in Kruskal coordinates \cite{wald2024general} is
	\begin{equation}
		\mathrm{d}s^2=-\frac{32M^3}{r}\mathrm{e}^{-r/2M}(\mathrm{d}T^2-\mathrm{d}X^2)+r^2\mathrm{d}\Omega^2,\label{1}
	\end{equation}
	where
	\begin{equation}
		\left(\frac{r}{2M}-1\right)\mathrm{e}^{r/2M}=X^2-T^2\equiv \zeta,\label{2}
	\end{equation}
	and $\mathrm{d}\Omega^2=\mathrm{d}\theta^2+\sin^2\theta\mathrm{d}\varphi^2$.
	The spacetime is defined within the domain $\zeta>-1$, where the boundary $\zeta=-1$ ($r=0$) corresponds to the physical singularity.
	It possesses a bifurcate horizon at $\zeta=0$ ($r=2M$), which separates the asymptotically flat untrapped (UT) region ($\zeta>0$) from the trapped (T) and anti-trapped (AT) regions, the latter two being characterized by $-1<\zeta<0$.
	The Killing vector field is given by $\xi^a=[X(\partial/\partial T)^a+T(\partial/\partial X)^a]/(4M)$.
	
	In the following, we assume that for the regularized spacetime the coordinates $\{T, X, \theta, \varphi\}$ remain globally valid, satisfying $-g_{TT}=g_{XX}\equiv A(T,X)>0$ and $r>0$, and that the Killing vector field is given by $\xi^a$.
	According to the Killing equation $\mathscr{L}_\xi g_{ab}=0$ \cite{wald2024general}, we have $\xi^a\nabla_a A=0$ and $\xi^a\nabla_a r=0$; therefore $A$ and $r$ should both be functions of $\zeta$.
	The two radial null geodesic tangent vectors are defined as
	$k_{\pm}^a\equiv (1/\sqrt{2A})[(\partial/\partial T)^a\pm(\partial/\partial X)^a]$.
	Their expansions $\theta_{k_{\pm}}=h^{ab}\nabla_a k_{\pm b}$ \cite{wald2024general,yang2021trapping} are computed using the induced metric $h_{ab}\equiv g_{ab}+k_{+a} k_{-b}+k_{-a} k_{+b}$ and are given by $\theta_{k_{\pm}}=\sqrt{2/A}(r^\prime/r)(-T\pm X)$.
	Here $r^\prime\equiv \mathrm{d}r/\mathrm{d}\zeta$; therefore, $r^\prime$ should be positive to ensure $\theta_{k_{\pm}}\leq 0\implies -T\pm X\leq 0$, which is consistent with the Schwarzschild case.
	Therefore, the Killing horizon remains described by $\zeta=0$.
	Together with the requirement that Eq.~\eqref{2} be regained in the limit $r\to\infty$ (which corresponds to $\zeta\to+\infty$), the spacetime is defined within the domain $\zeta > \zeta_i$, where the lower bound $\zeta_i<0$ is determined by the specific regularization scheme.
	Since $r^\prime>0$, as $\zeta\to \zeta_i$, we have $r\to L=\text{const}>0$, where $r=L$ is expected to be infinity of the spacetime.
	
	Since in Eq.~\eqref{1}, we have $A=8M/\zeta^\prime$ where $\zeta^\prime\equiv \mathrm{d}\zeta/\mathrm{d}r$, if one impose $A=8M/\zeta^\prime$, one will obtain a contradiction with that $r=L$ is infinity.
	When $\zeta<0$, define $X\equiv \sqrt{-\zeta}\sinh [t/(4M)]$ and $T\equiv \sqrt{-\zeta}\cosh[t/(4M)]$, one have $-A(\mathrm{d}T^2-\mathrm{d}X^2)$ equals to
	\begin{equation}
		-\frac{A\zeta}{16M^2}\mathrm{d}t^2+\frac{A\zeta^{\prime 2}}{4\zeta}\mathrm{d}r^2,\label{3}
	\end{equation}
	therefore $-g_{tt}=g^{-1}_{rr}=\zeta/(2M\zeta^\prime)$.
	Clearly, we have $g_{tt} g_{rr}=-1$, since $\xi^a=(\partial/\partial t)^a$ is the Killing vector field, for radial null geodesics, we have $p_t\equiv -g_{tt}dt/d\lambda=\text{const}$ and its affine parameter $\lambda|_{L}=\lambda|_{r_0}\pm\int_{r_0}^{L}\sqrt{-g_{tt}g_{rr}}\mathrm{d}r/p_t$, therefore $r=L$ cannot be infinity since $\lambda|_{L}$ is finite.
	Therefore, a general regularization is applied to both $A$ and $r$ via Eq.~\eqref{2}.
	Together with the condition that the spacetime could reduce to the Schwarzschild black hole in an appropriate limit and is asymtotically flat, such regular black holes are thoretically widely existed.
	
	Define null coordinates $u\equiv T-X$ and $v\equiv T+X$.
	Since $\zeta = -uv$, there are two different classes of the regularized spacetime:
	(i) The infinity of the spacetime inside the horizon is described by $\zeta_i=-\infty$.
	Since at infinity we have $r=L$, the infinity cannot be a conformal boundary; it consists of ideal boundaries $(u = \text{const} > 0, v = +\infty)$ and $(u = +\infty, v = \text{const} > 0)$ together with the ideal point $(u = +\infty, v = +\infty)$.
	Analogous to null infinity, two different types of radial null geodesics cannot share the same ideal endpoint; hence we refer to it as null-type infinity.
	(ii) The infinity of the spacetime inside the horizon is described by $\zeta_i = \text{const} < 0$. Since the surface $\zeta = \text{const} < 0$ is spacelike, similar to spacelike infinity, two different types of radial null geodesics can have the same ideal endpoint; therefore we refer to it as spacelike-type infinity.
	
	It is shown that any spherically symmetric metric of the form in Eq.~\eqref{1} is an exact solution to a coupled scalar–electromagnetic system \cite{liang2025globally}
	\begin{align}
		S = \int_M \epsilon \Big[ & R- p_1 \nabla^a \phi_1 \nabla_a \phi_1- p_2 \nabla^a \phi_2 \nabla_a \phi_2 \notag\\
		& - p_3 \nabla^a \phi_1 \nabla_a \phi_2- p_4 F \Big]
	\end{align}
	(with $c=G=4\pi\varepsilon_0=1$), consisting of two scalar fields and a magnetic field $F_{ab}=g\sin\theta(\mathrm{d}\theta)_a\wedge(\mathrm{d}\varphi)_b$, where $F = F_{ab}F^{ab} = 2g^2/r^4$ and $g$ is a constant \cite{liang2025globally}.
	The functions $p_1,p_2,p_3$ and $ p_4$ all depend on the scalar fields $\phi_1$ and $\phi_2$.
	By the reparamatrization $\phi_1=X,\phi_2=T$, the solution can be solved as  
	\begin{gather}
		p_1=G_{TT}-\frac{A}{r^2}G_{\theta\theta},\ p_2=G_{XX}+\frac{A}{r^2}G_{\theta\theta},\\
		p_3=2G_{TX},\ p_4=\frac{1}{A F}\left(G_{TT}-G_{XX}\right).
	\end{gather}

	Let $\mathscr{K}$ denote the Killing horizon associated with the Killing vector field $\xi^a$, and $\zeta=0$ when $r=r_{\mathscr{K}}$.
	Since the surface gravity $\kappa$ is defined by $\xi^b\nabla_b \xi^a|_{\mathscr{K}}=\kappa \xi^a|_{\mathscr{K}}$, according to Eq.~\eqref{3}, we have
	\begin{equation}
		\kappa=-\frac{\partial_r g_{tt}}{2\sqrt{-g_{tt}g_{rr}}}\bigg|_{\mathscr{K}}=\frac{1}{4M},
	\end{equation}
	where we have used the condition that $A\neq0,\zeta|_{\mathscr{K}}=0$ and $\zeta^\prime>0$.
	Therefore, $\kappa$ is invariant under a general regularization, and the Hawking temperature $T=\kappa/(2\pi)$ coincides with that of the Schwarzschild black hole.

	In the following, two concrete examples---one corresponding to null‑type infinity and the other to spacelike‑type infinity inside the horizon---are presented and examined.

	\textit{$\zeta_i=-\infty$.}---
	Since such regular black holes are widely present, to provide a concrete example in this study we base our selection on two principles: (i) simplicity; (ii) attempting to satisfy the energy conditions.
	Therefore, we assume that $A$ is invariant (since as $r\to L$, $A$ in Eq.~\eqref{1} is finite and nonzero), and impose
	\begin{equation}
		\left(\frac{r}{2M}-1\right)\mathrm{e}^{r/2M}[1+f(r)]=\zeta,\label{6}
	\end{equation}
	so that $r=2M$ remains the horizon radius and $L < 2M$, with a smooth and analytic function $f$ such that $f = o(1/r)$ as $r\to\infty$.
	Then, we should ensure that the spacetime satisfies the energy conditions at least in the limit $r \to \infty$.
	This requirement depends on the decay rate of $|f|$ as $r \to \infty$.
	For $\zeta>0$, we define $X\equiv \sqrt{\zeta}\cosh [t/(4M)]$ and $T\equiv \sqrt{\zeta}\sinh[t/(4M)]$.
	The metric then reads
	\begin{equation}
		\mathrm{d}s^2=\frac{2M\zeta}{r}\mathrm{e}^{-r/2M}\left(-\mathrm{d}t^2+\frac{4M^2\zeta^{\prime 2}}{\zeta^2}\mathrm{d}r^2\right)+r^2\mathrm{d}\Omega^2.\label{nulltype}
	\end{equation}
	The analysis of energy conditions is given in the Appendix and shows that, as $r\to\infty$, at most the WEC and SEC can be satisfied.
	In the following, we adopt class III with $f=w(r) e^{-r/2M}$ to give an example, where $w$ is a sub-exponentially varying function and $\lim_{r\to\infty}rw^\prime/w<-1$.
	From the energy conditions, it must be that $w>0$.
	According to Eq.~\eqref{6}, as $\zeta\to-\infty$ ($r\to L$), since $L<2M$, it must be that $w(r)$ diverges to $+\infty$ as $r\to L$.
	The natural choice is that $w=(r/L-1)^{-2}$ (therefore $\lim_{r\to\infty}rw^\prime/w=-2<-1$).
	The monotonicity of $r(\zeta)$ cannot be examined analytically; however, for sufficiently small $L$ (as expected), $r^\prime$ is positive.
	
\begin{figure}[t]      
		\centering             
		\includegraphics[width=0.45\textwidth]{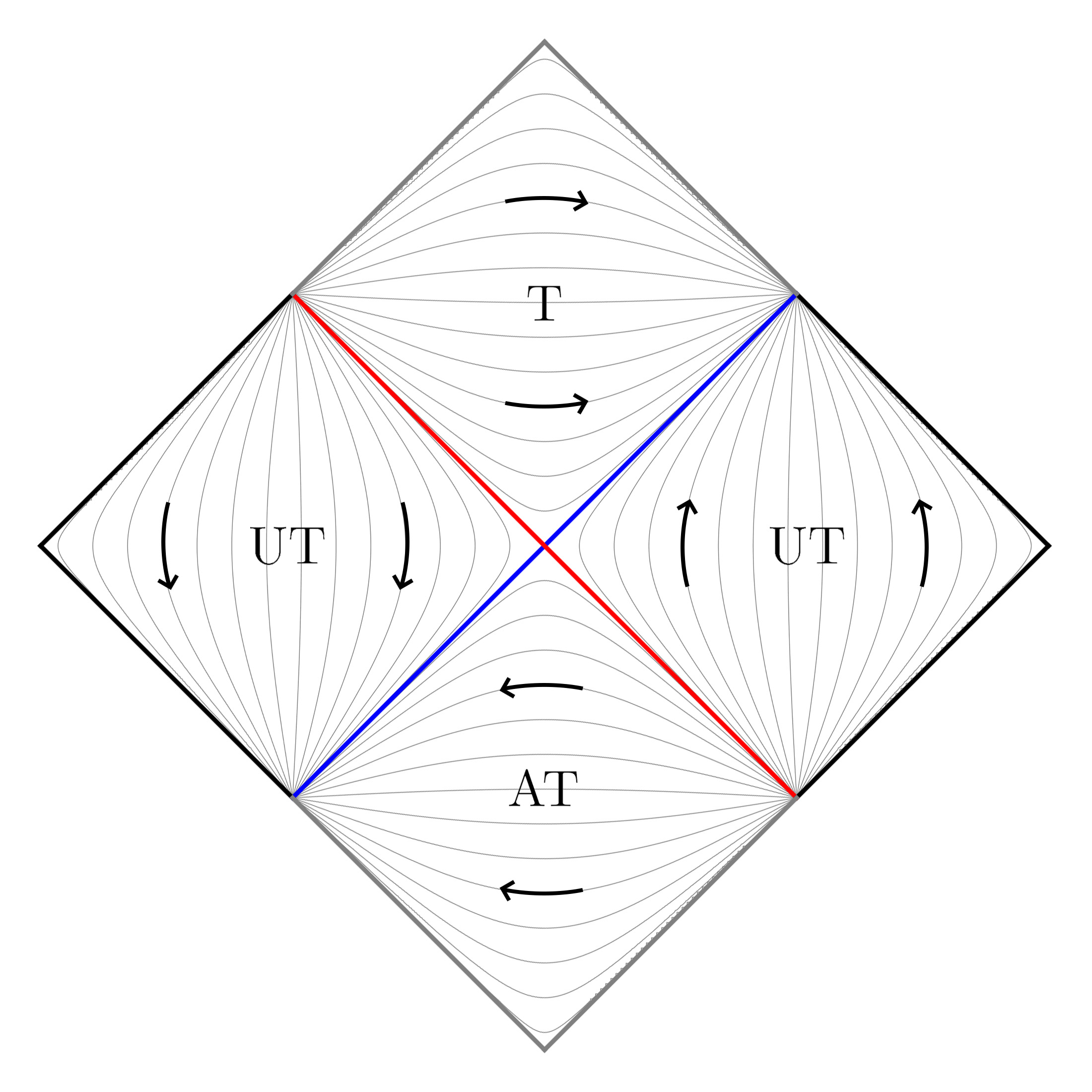}
		\caption{The causal diagram for the regularized spacetime with $\zeta_i=-\infty$.
			Regions: $\mathrm{T}$ (trapped), $\mathrm{AT}$ (anti-trapped), $\mathrm{UT}$ (untrapped).
			Blue and red lines indicate marginally trapping horizons defined by $k^a_+$ and $k^a_-$ respectively; four thick black lines denote the null infinities described by $r=\infty$; thin gray lines denote constant $r$ surfaces (larger $r$ closer to null infinity); four thick gray lines denote the null-type infinities described by $r=L$.
			Black arrows denote the Killing vector field $(\partial/\partial t)^a$.}
		\label{f1}    
	\end{figure}

	Then it should be verified that $r=L$ indeed corresponds to infinity.
	Solving the geodesic equations \cite{chandrasekhar1998mathematical} in coordinates $\{t,r,\theta,\varphi\}$, one finds that for timelike geodesics and non‑radial null geodesics, $u\to+\infty$ and $v\to+\infty$.
	For radial null geodesics, we have either $v=\text{const}>0$ and $u\to+\infty$, or $u=\text{const}>0$ and $v\to+\infty$.
	The Appendix provides the asymptotic behavior of curvature invariants and the energy-momentum tensor.
	Field sources are too lengthy to present here.
	FIG~\ref{f1} shows the causal diagram of the spacetime.
	Since $r_{\mathscr{K}}=2M$, the entropy $S$ is $4M^2\pi $ and the heat capacity $C$ is $-8\pi M^2$ coincides with Schwarzschild black hole.
	According to the Generalized Uncertainty Principle \cite{adler2001generalized}, $M$ cannot evaporate to zero; therefore $T$ approaches a finite maximum and $C\to 0^-$ as $M$ evaporates to its finite minimum.
	This is consistent with the existence of a sufficiently small $L$ that prevents $M\to0$.
	
\begin{figure}[htbp]      
	\centering             
	\includegraphics[width=0.45\textwidth]{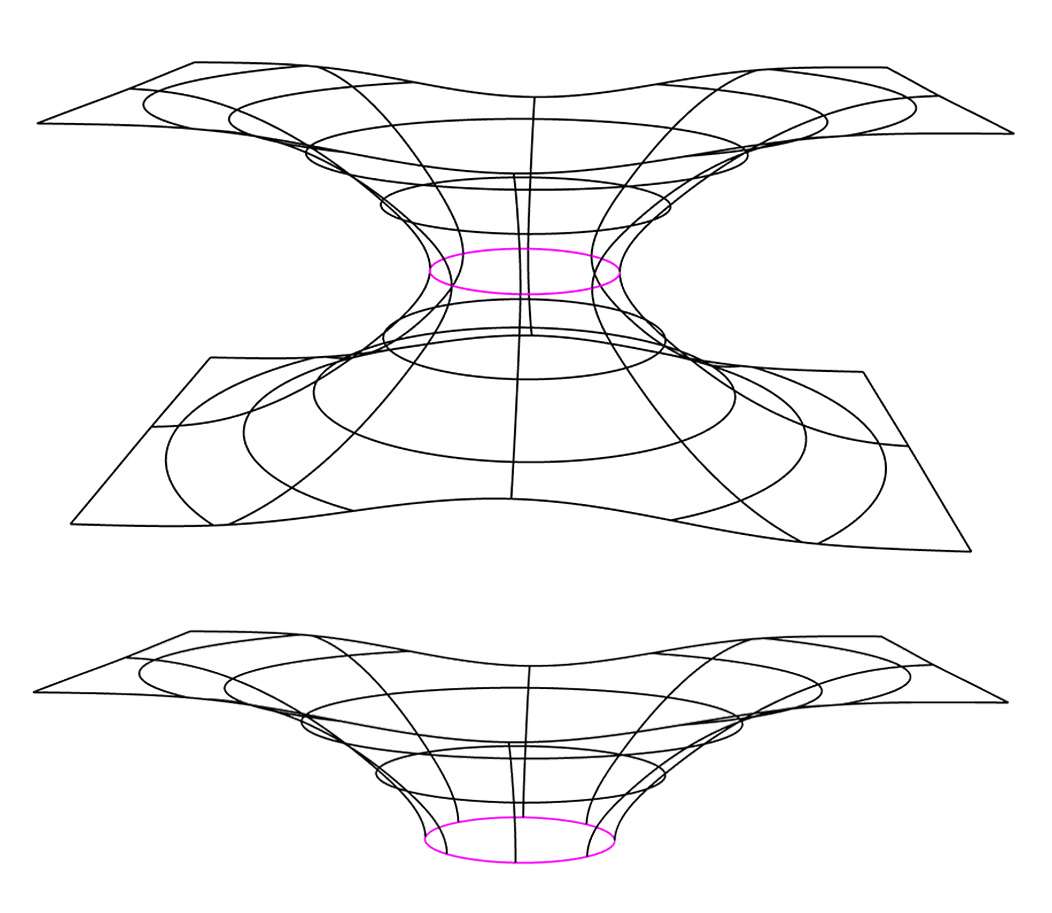}
	\caption{The constant-$T$ spacelike slice. }
	\label{f3}    
\end{figure}

	The remaining problem is that such a spacetime requires two asymptotically flat UT regions, meaning two universes, which seems less realistic.
	A spacelike slice ($T$ is constant) is shown in the upper panel in FIG.~\ref{f3}.
	To obtain a more realistic spacetime, the spacetime manifold is constructed as a quotient space under the discrete symmetry $X\to-X$; therefore, a spacelike slice ($T$ is constant) is shown in the lower panel in FIG.~\ref{f3}.
	The orange circle is just the reflection of the minimal radius circle on the spacelike slice.
	For $T\to\infty$, its radius reaches $L$, meaning it is the minimal describable radius in quantum gravity.
	In this case, the curves that reach $X=0$ experience a reflection, which causes them not even to be $C^1$.
	The explanation of such a phenomenon should rely on further study in quantum gravity.

\begin{figure}[t]      
	\centering             
	\includegraphics[width=0.45\textwidth]{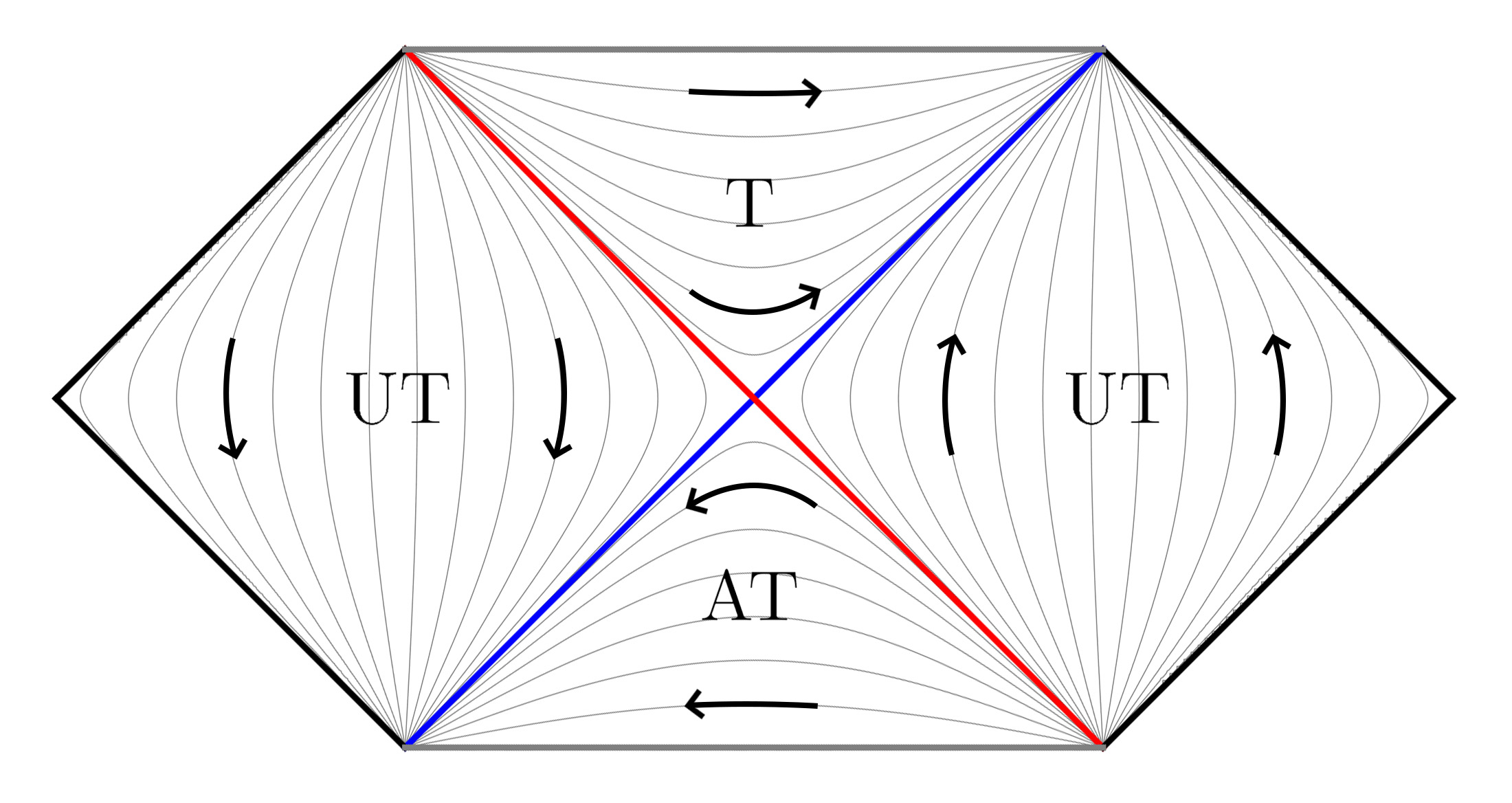}
	\caption{The causal diagram for the regularized spacetime with $\zeta_i\neq-\infty$.}
	\label{f2}    
\end{figure}

	\textit{$\zeta_i\neq-\infty$.}---
	In this case, since $\zeta_i$ is a negative constant, a point on the boundary is defined by constant values of $u$ and $v$, or constant values of $T$ and $X$.
	If the boundary is infinity, then $A$ must diverge there.
	Hence, the necessary regularization is to regularize $A$.
	For simplicity, we leave $r$ (that is, the relation given by Eq.~\eqref{2}) invariant.
	Then we suppose $A$ is given by
	\begin{equation}
		\frac{32M^3}{r}\mathrm{e}^{-r/2M}[1+h(r)],
	\end{equation}
	with a smooth and analytic function $h$ such that $\lim_{r\to\infty}h=0$.
	The metric then reads
	\begin{equation}
		\mathrm{d}s^2=\left(1+h\right)\left(-A_{Schw}\mathrm{d}t^2+A_{Schw}^{-1}\mathrm{d}r^2\right)+r^2\mathrm{d}\Omega^2,\label{spacetype}
	\end{equation}
	where $A_{Schw}\equiv1-2M/r$, so it should also be that $h = o(1/r)$ as $r\to\infty$.
	The metric is much simpler than that given in Eq.~\eqref{nulltype}, while the analysis of energy conditions in the Appendix shows that even the NEC is violated as $r\to\infty$; therefore, the WEC, DEC, and SEC are also violated.
	Even so, since Eq.~\eqref{spacetype} is just a simple assumption, a generalized regularization which contains the regularization of both $A$ and $r$ may satisfy some energy conditions.
	To see the mechanism that makes $\zeta=\zeta_i$ a regular infinity, we proceed.
	
	Because $A$ must diverge at $r=L$, we take $h$ to be analogous to $w$ and set $h=(r/L-1)^{k}$, where $k<0$.
	Solving the geodesic equations as $r\to L$, one can verify that if $k\leq-2$, then $r=L$ indeed corresponds to infinity, where both $u$ and $v$ are constant for a point on that boundary.
	Taking $k=-2$ as an example, the asymptotic behavior and field sources are as described in the Appendix; the causal diagram is shown in FIG.~\ref{f2}.
	Thermodynamics is the same as for $\zeta_i=-\infty$.
	The realistic spacetime is constructed as a quotient space under the discrete symmetry $X\to-X$.
	
	\textit{Conclusions and discussion.}---
	In summary, we have established the asymptotic throat as a geometrically inevitable mechanism for singularity resolution in black hole interiors. 
	By imposing a fundamental minimal length scale, the classical singularity is naturally replaced by future infinity, elevating the asymptotic throat from a heuristic concept to a rigorous causal boundary.
	We have provided a general framework to explicitly regularize the Schwarzschild spacetime and constructed two concrete examples, corresponding to null-type and spacelike-type infinities, whose physical sources are identified within a scalar-electromagnetic theory.
	The horizon structure, surface gravity, and Hawking temperature remain unaltered.
	
	Several directions warrant further investigation.
	A more general regularization---modifying both the metric function $A$ and the radial coordinate relation $r$---may yield improved compliance with energy conditions. 
	The framework extends naturally to other black hole spacetimes; for instance, in the Reissner-Nordström case, the minimal radius can be chosen to satisfy $r_- < L < r_+$, thereby excising the Cauchy horizon and avoiding the associated mass inflation instability \cite{chandrasekhar1998mathematical}.
	Based on the identified field sources, linear perturbation analysis can be pursued to assess the stability of these regularized geometries.
	Finally, the dynamical formation of such structures and their stability under backreaction remain crucial open questions whose resolution likely requires a full quantum gravitational treatment.
	The asymptotic throat thus provides a pristine classical bedrock for these future investigations into the quantum nature of black hole interiors.

	\appendix
	\section{Appendix}
	\label{app:proof}
	
	Under the assumption of Eq.~\eqref{6}, we focus on the case where (i) $f^{\prime\prime}/f^\prime$ has a limit (finite or infinite) as $r\to\infty$, (ii) $\lim_{r\to\infty} f'(r) = 0$, and (iii) $ f^{\prime 2}/f^{\prime\prime}$ and $ f^{\prime}/(r^2 f^{\prime\prime})$ have a limit as $r\to\infty$.
	From (i) and (ii), there are three different classes: (I) $\lim_{r\to\infty} |f^\prime/f|=\lim_{r\to\infty} |f^{\prime\prime}/f^\prime|=0$; (II) $\lim_{r\to\infty} |f^\prime/f|=\lim_{r\to\infty} |f^{\prime\prime}/f^\prime|=\infty$; (III) $\lim_{r\to\infty} |f^\prime/f|=\lim_{r\to\infty} |f^{\prime\prime}/f^\prime|=\beta$ (where $\beta$ is a positive constant).
	From (iii), it can be verified that $\lim_{r\to\infty} f^{\prime 2}/f^{\prime\prime}=0$ and $\lim_{r\to\infty} f^{\prime}/(r^2 f^{\prime\prime})=0$.
	
	For class (I), we have (keeping only the leading term)
	\begin{equation}
		\rho\to \frac{f+r f^\prime}{r^2},\ P_1\to \frac{-f+r f^\prime}{r^2},\ P_2=P_3\to  \frac{f^{\prime\prime}}{2}
	\end{equation}
	as $r\to+\infty$, where $\rho\equiv -G_{t}^{\; t} ,P_1\equiv G_{r}^{\; r},P_2\equiv G_{\theta} ^{\; \theta}=P_3\equiv G_{\varphi} ^{\; \varphi}$ and $G_{ab}$ is the Einstein tensor.
	Clearly, the null energy condition (NEC) $\rho+P_i>0$ leads to $f^\prime>0$ and $(f+rf^\prime)/r^2+f^{\prime\prime}/2>0$ (the latter is nonzero since $f=o(1/r)$ as $r\to\infty$).
	The first condition leads to $f\to0^-$ as $r\to\infty$.
	Define $g = r f$; therefore $g\to 0^-$ as $r\to\infty$.
	The latter condition leads to $g^{\prime\prime}>-2g/r^2>0$; then $g$ is a convex function and hence non-increasing.
	This contradicts $g\to 0^-$ as $r\to\infty$.
	Hence, even the NEC cannot be satisfied.
	Therefore, the weak energy condition (WEC), dominant energy condition (DEC) and strong energy condition (SEC) also cannot be satisfied.
	
	For class (II), we have
	\begin{equation}
		\rho\to \frac{4Mf^{\prime\prime}}{r},\ P_1\to \frac{f^\prime}{r},\ P_2=P_3\to \frac{f^{\prime\prime}}{2}
	\end{equation}
	as $r\to\infty$.
	The satisfaction of WEC implies $f^{\prime\prime}>0$, since $\rho+P_1\approx \rho$ and $\rho+P_2\approx P_2$.
	The SEC is satisfied if $\rho+P_1+2P_2\approx f^{\prime\prime}>0$.
	The DEC would require $\rho-P_1\approx\rho>0\implies f^{\prime\prime}>0$ and $\rho-P_2\approx-P_2>0\implies f^{\prime\prime}<0$.
	These two requirements are incompatible; hence DEC cannot be satisfied.
	
	For class (III), we have
	\begin{equation}
		\rho\to \frac{f^\prime+4Mf^{\prime\prime}}{r},\ P_1\to \frac{f^\prime}{r},\ P_2\to \frac{f^{\prime\prime}}{2}
	\end{equation}
	as $r\to\infty$.
	Generally, since $\rho+P_2\approx P_2>0\implies f^{\prime\prime}>0$ and $\rho-P_2\approx -P_2>0\implies f^{\prime\prime}<0$, WEC and DEC cannot be satisfied simultaneously, and SEC is satisfied if $\rho+P_1+2P_2\approx2P_2>0\implies f^{\prime\prime}>0$.
	Without loss of generality, $f$ can be parameterized as $f(r)=w(r)\mathrm{e}^{-r/a}$, where $w(r)$ is a sub-exponentially varying function and $a$ is a positive constant.
	Hence, $\lim_{r\to\infty}w^\prime/w=0$.
	Suppose also that $\lim_{r\to\infty}w^{\prime\prime}/w^\prime=0$.
	This case can be further divided into many classes; we therefore omit them and set $a=2M$ as a simple example.
	In this case, we have $\rho\to \mathrm{e}^{-r/2M}w/(2Mr),P_1\to -\mathrm{e}^{-r/2M}w/(2Mr)$; therefore $\rho>0\implies w>0$.
	Since the leading term of $\rho$ plus the leading term of $P_1$ vanishes, define $k(r)=rw^\prime/w$.
	Then $\rho+P_1>0$ implies $\lim_{r\to\infty}w(1+k)<0$ (we omit the case $\lim_{r\to\infty}k=-1$, as it would require an analysis of the asymptotic behaviour of $r(1+k)$ and would lead to even more classes).
	Therefore, if $\lim_{r\to\infty}k>-1$, then $\rho+P_1>0$ implies $w<0$, but $f^{\prime\prime}>0\implies w>0$; consequently the NEC cannot be satisfied.
	If $\lim_{r\to\infty}k<-1$, then $\rho+P_1>0$ implies $w>0$, and thus both WEC and SEC can be satisfied.

	By setting $f=(r/L-1)^{-2} e^{-r/2M}$, the curvature invariants $R$, $R_{ab}R^{ab}$ and $R_{abcd}R^{abcd}$ have the following asymptotic behavior, respectively, as $r\to\infty$:
	\begin{equation}
		-\frac{L^2\mathrm{e}^{-r/(2M)}}{4M^2r^2},\  \frac{L^4\mathrm{e}^{-r/M}}{32M^4r^4},\  \frac{48M^2}{r^6},
	\end{equation}
	and the following asymptotic behavior, respectively, as $r\to L$:
	\begin{equation}
		\frac{2}{L^2} ,\ \frac{2}{L^4},\ \frac{4}{L^4}.
	\end{equation}
	The energy density $\rho$ and pressures $P_1,P_2$ (defined in the Appendix) have the following asymptotic behavior, respectively, as $r\to\infty$:
	\begin{equation}
		\frac{L^2\mathrm{e}^{-r/(2M)}}{2Mr^3},\  -\frac{L^2\mathrm{e}^{-r/(2M)}}{2Mr^3},\ \frac{L^2\mathrm{e}^{-r/(2M)}}{8M^2r^2},
	\end{equation}
	and the following asymptotic behavior, respectively, as $r\to L$ (when $r<2M$, note that $\rho\equiv -G_{r}^{\; r}$ and $P_1\equiv G_{t}^{\; t}$):
	\begin{equation}
		\frac{1}{L^2} ,\ -\frac{1}{L^2},\ \frac{\mathrm{e}^{L/(2M)}(L-r)^3}{64M^3L^2}.
	\end{equation}
	
	Under the assumption of Eq.~\eqref{spacetype}, we also focus on the three requirements of $h$ similar to $f$.
	Therefore, there are also three classes, and we do not list them again here.
	For class (I), we have 
	\begin{equation}
		\rho\to \frac{h+r h^\prime}{r^2},\ P_1\to \frac{-h+r h^\prime}{r^2},\ P_2=P_3\to  \frac{h^{\prime\prime}}{2}
	\end{equation}
	as $r\to+\infty$.
	Similar to the case of $f$, the NEC cannot be satisfied.
	For class (II) and (III), we have 
	\begin{equation}
		\rho\to \frac{h^\prime}{r},\ P_1\to \frac{h^\prime}{r},\ P_2=P_3\to  \frac{h^{\prime\prime}}{2}
	\end{equation}
	as $r\to+\infty$.
	Then, $\rho+P_1>0$ implies $h^\prime>0$, and $\rho+P_2>0$ implies $h^{\prime\prime}>0$; but they cannot be satisfied simultaneously since $h\to 0$ as $r\to\infty$.
	Therefore, NEC cannot be satisfied.
	
	By setting $h=(r/L-1)^{-2}$, the curvature invariants $R$, $R_{ab}R^{ab}$ and $R_{abcd}R^{abcd}$ have the following asymptotic behavior, respectively, as $r\to\infty$:
	\begin{equation}
		-\frac{4L^2}{r^4},\  \frac{28 L^4}{r^8},\  \frac{48M^2}{r^6},
	\end{equation}
	and the following asymptotic behavior, respectively, as $r\to L$:
	\begin{equation}
		\frac{4M}{L^3} ,\ \frac{4[(M-L)^2+M^2]}{L^6},\ \frac{8[(M-L)^2+M^2]}{L^6}.
	\end{equation}
	The energy density $\rho$ and pressures $P_1,P_2$ have the following asymptotic behavior, respectively, as $r\to\infty$:
	\begin{equation}
		-\frac{L^2}{r^4},\  -\frac{3L^2}{r^4},\ \frac{3L^2}{r^4},
	\end{equation}
	and the following asymptotic behavior, respectively, as $r\to L$ (when $r<2M$, note that $\rho\equiv -G_{r}^{\; r}$ and $P_1\equiv G_{t}^{\; t}$):
	\begin{equation}
		\frac{1}{L^2} ,\ -\frac{1}{L^2},\ \frac{L-2M}{L^3}.
	\end{equation}
	
\bibliography{AT}

\begin{thebibliography}{30}%
\makeatletter
\providecommand \@ifxundefined [1]{%
 \@ifx{#1\undefined}
}%
\providecommand \@ifnum [1]{%
 \ifnum #1\expandafter \@firstoftwo
 \else \expandafter \@secondoftwo
 \fi
}%
\providecommand \@ifx [1]{%
 \ifx #1\expandafter \@firstoftwo
 \else \expandafter \@secondoftwo
 \fi
}%
\providecommand \natexlab [1]{#1}%
\providecommand \enquote  [1]{``#1''}%
\providecommand \bibnamefont  [1]{#1}%
\providecommand \bibfnamefont [1]{#1}%
\providecommand \citenamefont [1]{#1}%
\providecommand \href@noop [0]{\@secondoftwo}%
\providecommand \href [0]{\begingroup \@sanitize@url \@href}%
\providecommand \@href[1]{\@@startlink{#1}\@@href}%
\providecommand \@@href[1]{\endgroup#1\@@endlink}%
\providecommand \@sanitize@url [0]{\catcode `\\12\catcode `\$12\catcode
  `\&12\catcode `\#12\catcode `\^12\catcode `\_12\catcode `\%12\relax}%
\providecommand \@@startlink[1]{}%
\providecommand \@@endlink[0]{}%
\providecommand \url  [0]{\begingroup\@sanitize@url \@url }%
\providecommand \@url [1]{\endgroup\@href {#1}{\urlprefix }}%
\providecommand \urlprefix  [0]{URL }%
\providecommand \Eprint [0]{\href }%
\providecommand \doibase [0]{https://doi.org/}%
\providecommand \selectlanguage [0]{\@gobble}%
\providecommand \bibinfo  [0]{\@secondoftwo}%
\providecommand \bibfield  [0]{\@secondoftwo}%
\providecommand \translation [1]{[#1]}%
\providecommand \BibitemOpen [0]{}%
\providecommand \bibitemStop [0]{}%
\providecommand \bibitemNoStop [0]{.\EOS\space}%
\providecommand \EOS [0]{\spacefactor3000\relax}%
\providecommand \BibitemShut  [1]{\csname bibitem#1\endcsname}%
\let\auto@bib@innerbib\@empty
\bibitem [{\citenamefont
  {Schwarzschild}(1916)}]{schwarzschild1916gravitationsfeld}%
  \BibitemOpen
  \bibfield  {author} {\bibinfo {author} {\bibfnamefont {K.}~\bibnamefont
  {Schwarzschild}},\ }\bibfield  {title} {\bibinfo {title} {{\"U}ber das
  gravitationsfeld einer kugel aus inkompressibler fl{\"u}ssigkeit nach der
  einsteinschen theorie},\ }\href@noop {} {\bibfield  {journal} {\bibinfo
  {journal} {Sitzungsberichte der k{\"o}niglich preu{\ss}ischen Akademie der
  Wissenschaften zu Berlin}\ ,\ \bibinfo {pages} {424}} (\bibinfo {year}
  {1916})}\BibitemShut {NoStop}%
\bibitem [{\citenamefont {Penrose}(1965)}]{penrose1965gravitational}%
  \BibitemOpen
  \bibfield  {author} {\bibinfo {author} {\bibfnamefont {R.}~\bibnamefont
  {Penrose}},\ }\bibfield  {title} {\bibinfo {title} {Gravitational collapse
  and space-time singularities},\ }\href@noop {} {\bibfield  {journal}
  {\bibinfo  {journal} {Physical Review Letters}\ }\textbf {\bibinfo {volume}
  {14}},\ \bibinfo {pages} {57} (\bibinfo {year} {1965})}\BibitemShut {NoStop}%
\bibitem [{\citenamefont {Hawking}(1967)}]{hawking1967occurrence}%
  \BibitemOpen
  \bibfield  {author} {\bibinfo {author} {\bibfnamefont {S.~W.}\ \bibnamefont
  {Hawking}},\ }\bibfield  {title} {\bibinfo {title} {The occurrence of
  singularities in cosmology. iii. causality and singularities},\ }\href@noop
  {} {\bibfield  {journal} {\bibinfo  {journal} {Proceedings of the Royal
  Society of London. Series A. Mathematical and Physical Sciences}\ }\textbf
  {\bibinfo {volume} {300}},\ \bibinfo {pages} {187} (\bibinfo {year}
  {1967})}\BibitemShut {NoStop}%
\bibitem [{\citenamefont {Hawking}\ and\ \citenamefont
  {Penrose}(1970)}]{hawking1970singularities}%
  \BibitemOpen
  \bibfield  {author} {\bibinfo {author} {\bibfnamefont {S.~W.}\ \bibnamefont
  {Hawking}}\ and\ \bibinfo {author} {\bibfnamefont {R.}~\bibnamefont
  {Penrose}},\ }\bibfield  {title} {\bibinfo {title} {The singularities of
  gravitational collapse and cosmology},\ }\href@noop {} {\bibfield  {journal}
  {\bibinfo  {journal} {Proceedings of the Royal Society of London. A.
  Mathematical and Physical Sciences}\ }\textbf {\bibinfo {volume} {314}},\
  \bibinfo {pages} {529} (\bibinfo {year} {1970})}\BibitemShut {NoStop}%
\bibitem [{\citenamefont {Bardeen}(1968)}]{bardeen1968non}%
  \BibitemOpen
  \bibfield  {author} {\bibinfo {author} {\bibfnamefont {J.}~\bibnamefont
  {Bardeen}},\ }\bibfield  {title} {\bibinfo {title} {Non-singular general
  relativistic gravitational collapse},\ }in\ \href@noop {} {\emph {\bibinfo
  {booktitle} {Proceedings of the 5th International Conference on Gravitation
  and the Theory of Relativity}}}\ (\bibinfo {year} {1968})\ p.~\bibinfo
  {pages} {87}\BibitemShut {NoStop}%
\bibitem [{\citenamefont {Ayon-Beato}\ and\ \citenamefont
  {Garcia}(1998)}]{ayon1998regular}%
  \BibitemOpen
  \bibfield  {author} {\bibinfo {author} {\bibfnamefont {E.}~\bibnamefont
  {Ayon-Beato}}\ and\ \bibinfo {author} {\bibfnamefont {A.}~\bibnamefont
  {Garcia}},\ }\bibfield  {title} {\bibinfo {title} {Regular black hole in
  general relativity coupled to nonlinear electrodynamics},\ }\href@noop {}
  {\bibfield  {journal} {\bibinfo  {journal} {Physical Review Letters}\
  }\textbf {\bibinfo {volume} {80}},\ \bibinfo {pages} {5056} (\bibinfo {year}
  {1998})}\BibitemShut {NoStop}%
\bibitem [{\citenamefont {Ayon-Beato}\ and\ \citenamefont
  {Garc{\i}a}(2000)}]{ayon2000bardeen}%
  \BibitemOpen
  \bibfield  {author} {\bibinfo {author} {\bibfnamefont {E.}~\bibnamefont
  {Ayon-Beato}}\ and\ \bibinfo {author} {\bibfnamefont {A.}~\bibnamefont
  {Garc{\i}a}},\ }\bibfield  {title} {\bibinfo {title} {The bardeen model as a
  nonlinear magnetic monopole},\ }\href@noop {} {\bibfield  {journal} {\bibinfo
   {journal} {Physics Letters B}\ }\textbf {\bibinfo {volume} {493}},\ \bibinfo
  {pages} {149} (\bibinfo {year} {2000})}\BibitemShut {NoStop}%
\bibitem [{\citenamefont {Bronnikov}(2001)}]{bronnikov2001regular}%
  \BibitemOpen
  \bibfield  {author} {\bibinfo {author} {\bibfnamefont {K.~A.}\ \bibnamefont
  {Bronnikov}},\ }\bibfield  {title} {\bibinfo {title} {Regular magnetic black
  holes and monopoles from nonlinear electrodynamics},\ }\href@noop {}
  {\bibfield  {journal} {\bibinfo  {journal} {Physical Review D}\ }\textbf
  {\bibinfo {volume} {63}},\ \bibinfo {pages} {044005} (\bibinfo {year}
  {2001})}\BibitemShut {NoStop}%
\bibitem [{\citenamefont {Burinskii}\ and\ \citenamefont
  {Hildebrandt}(2002)}]{burinskii2002new}%
  \BibitemOpen
  \bibfield  {author} {\bibinfo {author} {\bibfnamefont {A.}~\bibnamefont
  {Burinskii}}\ and\ \bibinfo {author} {\bibfnamefont {S.~R.}\ \bibnamefont
  {Hildebrandt}},\ }\bibfield  {title} {\bibinfo {title} {New type of regular
  black holes and particlelike solutions from nonlinear electrodynamics},\
  }\href@noop {} {\bibfield  {journal} {\bibinfo  {journal} {Physical Review
  D}\ }\textbf {\bibinfo {volume} {65}},\ \bibinfo {pages} {104017} (\bibinfo
  {year} {2002})}\BibitemShut {NoStop}%
\bibitem [{\citenamefont {Fan}\ and\ \citenamefont
  {Wang}(2016)}]{fan2016construction}%
  \BibitemOpen
  \bibfield  {author} {\bibinfo {author} {\bibfnamefont {Z.-Y.}\ \bibnamefont
  {Fan}}\ and\ \bibinfo {author} {\bibfnamefont {X.}~\bibnamefont {Wang}},\
  }\bibfield  {title} {\bibinfo {title} {Construction of regular black holes in
  general relativity},\ }\href@noop {} {\bibfield  {journal} {\bibinfo
  {journal} {Physical Review D}\ }\textbf {\bibinfo {volume} {94}},\ \bibinfo
  {pages} {124027} (\bibinfo {year} {2016})}\BibitemShut {NoStop}%
\bibitem [{\citenamefont {Bronnikov}(2023)}]{bronnikov2023regular}%
  \BibitemOpen
  \bibfield  {author} {\bibinfo {author} {\bibfnamefont {K.~A.}\ \bibnamefont
  {Bronnikov}},\ }\bibfield  {title} {\bibinfo {title} {Regular black holes
  sourced by nonlinear electrodynamics},\ }in\ \href@noop {} {\emph {\bibinfo
  {booktitle} {Regular Black Holes: Towards a New Paradigm of Gravitational
  Collapse}}}\ (\bibinfo  {publisher} {Springer},\ \bibinfo {year} {2023})\
  pp.\ \bibinfo {pages} {37--67}\BibitemShut {NoStop}%
\bibitem [{\citenamefont {Simpson}\ and\ \citenamefont
  {Visser}(2019)}]{simpson2019black}%
  \BibitemOpen
  \bibfield  {author} {\bibinfo {author} {\bibfnamefont {A.}~\bibnamefont
  {Simpson}}\ and\ \bibinfo {author} {\bibfnamefont {M.}~\bibnamefont
  {Visser}},\ }\bibfield  {title} {\bibinfo {title} {Black-bounce to
  traversable wormhole},\ }\href@noop {} {\bibfield  {journal} {\bibinfo
  {journal} {Journal of Cosmology and Astroparticle Physics}\ }\textbf
  {\bibinfo {volume} {2019}}\bibinfo  {number} { (02)},\ \bibinfo {pages}
  {042}}\BibitemShut {NoStop}%
\bibitem [{\citenamefont {Lobo}\ \emph {et~al.}(2021)\citenamefont {Lobo},
  \citenamefont {Rodrigues}, \citenamefont {Silva}, \citenamefont {Simpson},\
  and\ \citenamefont {Visser}}]{lobo2021novel}%
  \BibitemOpen
\bibfield  {number} {  }\bibfield  {author} {\bibinfo {author} {\bibfnamefont
  {F.~S.}\ \bibnamefont {Lobo}}, \bibinfo {author} {\bibfnamefont {M.~E.}\
  \bibnamefont {Rodrigues}}, \bibinfo {author} {\bibfnamefont {M.~V. d.~S.}\
  \bibnamefont {Silva}}, \bibinfo {author} {\bibfnamefont {A.}~\bibnamefont
  {Simpson}},\ and\ \bibinfo {author} {\bibfnamefont {M.}~\bibnamefont
  {Visser}},\ }\bibfield  {title} {\bibinfo {title} {Novel black-bounce
  spacetimes: wormholes, regularity, energy conditions, and causal structure},\
  }\href@noop {} {\bibfield  {journal} {\bibinfo  {journal} {Physical Review
  D}\ }\textbf {\bibinfo {volume} {103}},\ \bibinfo {pages} {084052} (\bibinfo
  {year} {2021})}\BibitemShut {NoStop}%
\bibitem [{\citenamefont {Bronnikov}(2022)}]{bronnikov2022black}%
  \BibitemOpen
  \bibfield  {author} {\bibinfo {author} {\bibfnamefont {K.}~\bibnamefont
  {Bronnikov}},\ }\bibfield  {title} {\bibinfo {title} {Black bounces,
  wormholes, and partly phantom scalar fields},\ }\href@noop {} {\bibfield
  {journal} {\bibinfo  {journal} {Physical Review D}\ }\textbf {\bibinfo
  {volume} {106}},\ \bibinfo {pages} {064029} (\bibinfo {year}
  {2022})}\BibitemShut {NoStop}%
\bibitem [{\citenamefont {Bronnikov}\ and\ \citenamefont
  {Walia}(2022)}]{bronnikov2022field}%
  \BibitemOpen
  \bibfield  {author} {\bibinfo {author} {\bibfnamefont {K.~A.}\ \bibnamefont
  {Bronnikov}}\ and\ \bibinfo {author} {\bibfnamefont {R.~K.}\ \bibnamefont
  {Walia}},\ }\bibfield  {title} {\bibinfo {title} {Field sources for
  simpson-visser spacetimes},\ }\href@noop {} {\bibfield  {journal} {\bibinfo
  {journal} {Physical Review D}\ }\textbf {\bibinfo {volume} {105}},\ \bibinfo
  {pages} {044039} (\bibinfo {year} {2022})}\BibitemShut {NoStop}%
\bibitem [{\citenamefont {Ca{\~n}ate}(2022)}]{canate2022black}%
  \BibitemOpen
  \bibfield  {author} {\bibinfo {author} {\bibfnamefont {P.}~\bibnamefont
  {Ca{\~n}ate}},\ }\bibfield  {title} {\bibinfo {title} {Black bounces as
  magnetically charged phantom regular black holes in einstein-nonlinear
  electrodynamics gravity coupled to a self-interacting scalar field},\
  }\href@noop {} {\bibfield  {journal} {\bibinfo  {journal} {Physical Review
  D}\ }\textbf {\bibinfo {volume} {106}},\ \bibinfo {pages} {024031} (\bibinfo
  {year} {2022})}\BibitemShut {NoStop}%
\bibitem [{\citenamefont {Rovelli}(2004)}]{rovelli2004quantum}%
  \BibitemOpen
  \bibfield  {author} {\bibinfo {author} {\bibfnamefont {C.}~\bibnamefont
  {Rovelli}},\ }\href@noop {} {\emph {\bibinfo {title} {Quantum gravity}}}\
  (\bibinfo  {publisher} {Cambridge university press},\ \bibinfo {year}
  {2004})\BibitemShut {NoStop}%
\bibitem [{\citenamefont {Liang}\ and\ \citenamefont
  {Li}(2025)}]{liang2025globally}%
  \BibitemOpen
  \bibfield  {author} {\bibinfo {author} {\bibfnamefont {Y.-b.}\ \bibnamefont
  {Liang}}\ and\ \bibinfo {author} {\bibfnamefont {H.-R.}\ \bibnamefont {Li}},\
  }\bibfield  {title} {\bibinfo {title} {The globally trapped future: A fate
  for black holes and wormholes},\ }\href@noop {} {\bibfield  {journal}
  {\bibinfo  {journal} {arXiv preprint arXiv:2511.20427}\ } (\bibinfo {year}
  {2025})}\BibitemShut {NoStop}%
\bibitem [{\citenamefont {Hayward}(1999)}]{hayward1999dynamic}%
  \BibitemOpen
  \bibfield  {author} {\bibinfo {author} {\bibfnamefont {S.~A.}\ \bibnamefont
  {Hayward}},\ }\bibfield  {title} {\bibinfo {title} {Dynamic wormholes},\
  }\href@noop {} {\bibfield  {journal} {\bibinfo  {journal} {International
  Journal of Modern Physics D}\ }\textbf {\bibinfo {volume} {8}},\ \bibinfo
  {pages} {373} (\bibinfo {year} {1999})}\BibitemShut {NoStop}%
\bibitem [{\citenamefont {Shinkai}\ and\ \citenamefont
  {Hayward}(2002)}]{shinkai2002fate}%
  \BibitemOpen
  \bibfield  {author} {\bibinfo {author} {\bibfnamefont {H.-a.}\ \bibnamefont
  {Shinkai}}\ and\ \bibinfo {author} {\bibfnamefont {S.~A.}\ \bibnamefont
  {Hayward}},\ }\bibfield  {title} {\bibinfo {title} {Fate of the first
  traversible wormhole: Black-hole collapse or inflationary expansion},\
  }\href@noop {} {\bibfield  {journal} {\bibinfo  {journal} {Physical Review
  D}\ }\textbf {\bibinfo {volume} {66}},\ \bibinfo {pages} {044005} (\bibinfo
  {year} {2002})}\BibitemShut {NoStop}%
\bibitem [{\citenamefont {Hayward}(2009)}]{hayward2009wormhole}%
  \BibitemOpen
  \bibfield  {author} {\bibinfo {author} {\bibfnamefont {S.~A.}\ \bibnamefont
  {Hayward}},\ }\bibfield  {title} {\bibinfo {title} {Wormhole dynamics in
  spherical symmetry},\ }\href@noop {} {\bibfield  {journal} {\bibinfo
  {journal} {Physical Review D}\ }\textbf {\bibinfo {volume} {79}},\ \bibinfo
  {pages} {124001} (\bibinfo {year} {2009})}\BibitemShut {NoStop}%
\bibitem [{\citenamefont {Hayward}(2006)}]{hayward2006formation}%
  \BibitemOpen
  \bibfield  {author} {\bibinfo {author} {\bibfnamefont {S.~A.}\ \bibnamefont
  {Hayward}},\ }\bibfield  {title} {\bibinfo {title} {Formation and evaporation
  of nonsingular black holes},\ }\href@noop {} {\bibfield  {journal} {\bibinfo
  {journal} {Physical Review Letters}\ }\textbf {\bibinfo {volume} {96}},\
  \bibinfo {pages} {031103} (\bibinfo {year} {2006})}\BibitemShut {NoStop}%
\bibitem [{\citenamefont {Simpson}\ \emph {et~al.}(2019)\citenamefont
  {Simpson}, \citenamefont {Martin-Moruno},\ and\ \citenamefont
  {Visser}}]{simpson2019vaidya}%
  \BibitemOpen
  \bibfield  {author} {\bibinfo {author} {\bibfnamefont {A.}~\bibnamefont
  {Simpson}}, \bibinfo {author} {\bibfnamefont {P.}~\bibnamefont
  {Martin-Moruno}},\ and\ \bibinfo {author} {\bibfnamefont {M.}~\bibnamefont
  {Visser}},\ }\bibfield  {title} {\bibinfo {title} {Vaidya spacetimes,
  black-bounces, and traversable wormholes},\ }\href@noop {} {\bibfield
  {journal} {\bibinfo  {journal} {Classical and Quantum Gravity}\ }\textbf
  {\bibinfo {volume} {36}},\ \bibinfo {pages} {145007} (\bibinfo {year}
  {2019})}\BibitemShut {NoStop}%
\bibitem [{\citenamefont {Lobo}\ \emph {et~al.}(2020)\citenamefont {Lobo},
  \citenamefont {Simpson},\ and\ \citenamefont {Visser}}]{lobo2020dynamic}%
  \BibitemOpen
  \bibfield  {author} {\bibinfo {author} {\bibfnamefont {F.~S.}\ \bibnamefont
  {Lobo}}, \bibinfo {author} {\bibfnamefont {A.}~\bibnamefont {Simpson}},\ and\
  \bibinfo {author} {\bibfnamefont {M.}~\bibnamefont {Visser}},\ }\bibfield
  {title} {\bibinfo {title} {Dynamic thin-shell black-bounce traversable
  wormholes},\ }\href@noop {} {\bibfield  {journal} {\bibinfo  {journal}
  {Physical Review D}\ }\textbf {\bibinfo {volume} {101}},\ \bibinfo {pages}
  {124035} (\bibinfo {year} {2020})}\BibitemShut {NoStop}%
\bibitem [{\citenamefont {Yang}\ and\ \citenamefont
  {Huang}(2021)}]{yang2021trapping}%
  \BibitemOpen
  \bibfield  {author} {\bibinfo {author} {\bibfnamefont {J.}~\bibnamefont
  {Yang}}\ and\ \bibinfo {author} {\bibfnamefont {H.}~\bibnamefont {Huang}},\
  }\bibfield  {title} {\bibinfo {title} {Trapping horizons of the evolving
  charged wormhole and black bounce},\ }\href@noop {} {\bibfield  {journal}
  {\bibinfo  {journal} {Physical Review D}\ }\textbf {\bibinfo {volume}
  {104}},\ \bibinfo {pages} {084005} (\bibinfo {year} {2021})}\BibitemShut
  {NoStop}%
\bibitem [{\citenamefont {Bronnikov}\ and\ \citenamefont
  {Fabris}(2006)}]{bronnikov2006regular}%
  \BibitemOpen
  \bibfield  {author} {\bibinfo {author} {\bibfnamefont {K.~A.}\ \bibnamefont
  {Bronnikov}}\ and\ \bibinfo {author} {\bibfnamefont {J.~C.}\ \bibnamefont
  {Fabris}},\ }\bibfield  {title} {\bibinfo {title} {Regular phantom black
  holes},\ }\href@noop {} {\bibfield  {journal} {\bibinfo  {journal} {Physical
  Review Letters}\ }\textbf {\bibinfo {volume} {96}},\ \bibinfo {pages}
  {251101} (\bibinfo {year} {2006})}\BibitemShut {NoStop}%
\bibitem [{\citenamefont {Carballo-Rubio}\ \emph {et~al.}(2020)\citenamefont
  {Carballo-Rubio}, \citenamefont {Di~Filippo}, \citenamefont {Liberati},\ and\
  \citenamefont {Visser}}]{carballo2020geodesically}%
  \BibitemOpen
  \bibfield  {author} {\bibinfo {author} {\bibfnamefont {R.}~\bibnamefont
  {Carballo-Rubio}}, \bibinfo {author} {\bibfnamefont {F.}~\bibnamefont
  {Di~Filippo}}, \bibinfo {author} {\bibfnamefont {S.}~\bibnamefont
  {Liberati}},\ and\ \bibinfo {author} {\bibfnamefont {M.}~\bibnamefont
  {Visser}},\ }\bibfield  {title} {\bibinfo {title} {Geodesically complete
  black holes},\ }\href@noop {} {\bibfield  {journal} {\bibinfo  {journal}
  {Physical Review D}\ }\textbf {\bibinfo {volume} {101}},\ \bibinfo {pages}
  {084047} (\bibinfo {year} {2020})}\BibitemShut {NoStop}%
\bibitem [{\citenamefont {Wald}(2024)}]{wald2024general}%
  \BibitemOpen
  \bibfield  {author} {\bibinfo {author} {\bibfnamefont {R.~M.}\ \bibnamefont
  {Wald}},\ }\href@noop {} {\emph {\bibinfo {title} {General relativity}}}\
  (\bibinfo  {publisher} {University of Chicago press},\ \bibinfo {year}
  {2024})\BibitemShut {NoStop}%
\bibitem [{\citenamefont
  {Chandrasekhar}(1998)}]{chandrasekhar1998mathematical}%
  \BibitemOpen
  \bibfield  {author} {\bibinfo {author} {\bibfnamefont {S.}~\bibnamefont
  {Chandrasekhar}},\ }\href@noop {} {\emph {\bibinfo {title} {The mathematical
  theory of black holes}}},\ Vol.~\bibinfo {volume} {69}\ (\bibinfo
  {publisher} {Oxford university press},\ \bibinfo {year} {1998})\BibitemShut
  {NoStop}%
\bibitem [{\citenamefont {Adler}\ \emph {et~al.}(2001)\citenamefont {Adler},
  \citenamefont {Chen},\ and\ \citenamefont {Santiago}}]{adler2001generalized}%
  \BibitemOpen
  \bibfield  {author} {\bibinfo {author} {\bibfnamefont {R.~J.}\ \bibnamefont
  {Adler}}, \bibinfo {author} {\bibfnamefont {P.}~\bibnamefont {Chen}},\ and\
  \bibinfo {author} {\bibfnamefont {D.~I.}\ \bibnamefont {Santiago}},\
  }\bibfield  {title} {\bibinfo {title} {The generalized uncertainty principle
  and black hole remnants},\ }\href@noop {} {\bibfield  {journal} {\bibinfo
  {journal} {General Relativity and Gravitation}\ }\textbf {\bibinfo {volume}
  {33}},\ \bibinfo {pages} {2101} (\bibinfo {year} {2001})}\BibitemShut
  {NoStop}%
\end{thebibliography}%

\end{document}